# Using Twitter Data to Determine Hurricane Category: An Experiment

**Songhui Yue, Jyothsna Kondari, Aibek Musave, Randy Smith**
**Department of Computer Science, University of Alabama**
**{syue2, jkondari, aibek, rsmith}@ua.edu**

**Songqing Yue**
**Department of Computer Science and Software Engineering, University of Wisconsin – Platteville**
**yues@uwplatt.edu**

**Abstract**

Social media posts contain an abundant amount of information about public opinion on major events, especially natural disasters such as hurricanes. Posts related to an event, are usually published by the users who live near the place of the event at the time of the event. Special correlation between the social media data and the events can be obtained using data mining approaches. This paper presents research work to find the mappings between social media data and the severity level of a disaster. Specifically, we have investigated the Twitter data posted during hurricanes Harvey and Irma, and attempted to find the correlation between the Twitter data of a specific area and the hurricane level in that area. Our experimental results indicate a positive correlation between them. We also present a method to predict the hurricane category for a specific area using relevant Twitter data.

## INTRODUCTION

Nowadays, disaster control and analysis using social media data is gaining great popularity (Kireyev et al. 2009; Reuter & Kaufhold 2017). When a disaster happens, people who live nearby use social media to express their opinions, emotions, and feelings, or to ask for help. Special correlation between the data and the events can be implicit, but valuable for situations like emergency response, and can be found through applying data mining techniques. For example, researchers have used different techniques to determine the occurrence of some types of disasters, including their places and tracks (Musaev et al. 2015).

Hurricane Harvey and Irma have attracted a lot of attention due to the scale and damage they brought to the affected areas. Hurricane Harvey was an intensely devastating Atlantic Hurricane that affected southeast Texas, USA. Harvey formed on August 17 and dissipated on September 3, 2017 (Hurricane Harvey Recap from https://weather.com 2017). It was the first major Hurricane that affected south Texas after Celia in 1970. It strengthened into a hurricane on August 24, with first landfall on San Jose Island, followed by another landfall near Rockport and Fulton, Texas the next day. Another Hurricane named Irma, which followed Harvey, was the strongest Atlantic Hurricane in terms of maximal constant winds, and formed on August 30, 2017 and dissipated on September 16, 2017. Irma caused adverse damage starting from the Caribbean to the Southeast United States, especially Florida (Hurricane Irma Recap from https://weather.com 2017). On September 10, Irma moved over the Florida Keys as category 4 hurricane, making landfall with a wind speed of 130 mph at Cudjoe Key, Florida (Landsea & Mello 2017). Irma made another landfall close to Macro Island, Florida with wind speed of 115 mph.

During the period of these two hurricanes, thousands of tweets were posted all over the world. We are interested in evaluating a hypothesis that the scale of a hurricane in a location can be determined based on a given tweet or a set of tweets from that location. This can be generalized into finding the relationship between tweets and the level of the Hurricane and using the relationship to predict the actual hurricane level. Given the above research questions, we have performed several experiments on a one-month (August 17, 2017 through September 16, 2017) geo-tagged Twitter data set, which contains 188,721 tweets, and every tweet has a hashtag "Harvey" or "Irma". After filtering out the irrelevant tweets, we build two data sets for Harvey and Irma. We annotate each of the tweets in both data sets with different categories depending on their location and the category of hurricane in that location. We then generate vectors for the data sets using Word2Vec (Mikolov et al. 2013) and Bag-of-Words (BOW) (Tsai 2012), after which, we apply different classification algorithms to evaluate the vectors.

The contributions of this paper are three-fold. First, by analyzing a current available Twitter data corpus, the hypothesis about a positive relationship between Twitter data and hurricane category is verified. Second, the methodology we use in this paper can be applied in future work and to different disaster type and an algorithm for predicting a specific area's hurricane category is presented. The third contribution is a release of a labeled data set of hurricane related tweets, which was generated over the course of the study.

## RELATED WORK

Ashktorab et al. introduce a Twitter-mining tool that extracts actionable information for disaster relief workers (Ashktorab et al. 2014). Those workers need to glean actionable knowledge from a large volume of tweets and status updates; however, tweets and status updates can be very noisy. The goal of their tool is to extract information relevant for first responders in real time and enable analysis after the disaster has occurred. They use a combination of classification, clustering and extraction methods to extract actionable information. Similarly, Cornelia et al. present an approach based on convolutional neural networks to identifying informative messages in social media streams during disaster events and show significant improvement in performance over models that use the "bag of words" and n-grams as features on several datasets of messages from flooding events (Cornelia et al. 2016).

Sakaki et al. use Twitter to investigate the real-time interaction of events such as earthquakes and propose an algorithm to monitor tweets and to detect a target event (Sakaki et al. 2010). They produce a probabilistic spatiotemporal model for the target event that can find the center and the trajectory of the event location, and their application can deliver much faster than the announcements that are broadcast by Japan Meteorological Agency (JMA) with high probability (96%). Musaev et al. in their work of evaluating a landslide detection system, LITMUS, use Twitter as the main source of social media data to help detect landslides. Their integrated approach of using physical sensors and social media sensors (including Twitter) enables the LITMUS system to have a better precision and accuracy of landslide detection (Musaev et al. 2015).

Twitter data is also used to predict stock market movements (Pagolu et al. 2016). Their work is based on the fact that social media can represent the public sentiment and opinion about current events, and especially Twitter has attracted a lot of attention from researchers for studying the public sentiment. To observe how much the changes in stock prices of a company, the rises and falls, are correlated with the public opinions being expressed in tweets about that company, they applied two different textual representations, Word2Vec and N-grams.

Individual awareness of the severity of a disaster could be a key reason why we can use a collection of individual's social media posts to predict the severity of a disaster. A related work about individual awareness and disaster explores the possibility in the context of drought, examining what variables determine individual awareness of drought and further exploring how drought awareness influences risk perception and policy preferences (Switzer & Vedlitz 2017). They analyze related data and the results indicate that drought severity is the most significant factor for drought awareness, while ideological and demographic variables also play a role.

## METHODOLOGY

In this section, we introduce the methodologies used in our research. The overall steps of our research are shown in Figure 1. We first briefly describe these steps and then, in the following subsections, we describe the setup of environmental issues and present more details for each step of Figure 1.

*Geo-tagged Tweets Corpus*: To achieve our research goal, the corpus of Twitter data needs to be extracted first. Geo-tagged tweets are the ones that have geo-coordinates or other type of location information in their meta-data. Using location information, we can identify the tweets published in any area of interest. In the subsection of "Twitter data corpus", we introduce the Twitter data set we use in this research.

*Location Estimation for Hurricane Category*: Referring to the information provided by wunderground (a commercial weather service) (https://wunderground.com Harvey/Irma 2017), we use the latitude and longitude information to record on the map with the locations' related hurricane information. In this step, we can find the related towns and cities for each hurricane category. For example, Rockport in Texas is of hurricane category 4.
*Data Filtering*: We search in the data corpus and find relevant tweets for each of the towns or cities identified for each hurricane category. The filters that we use include: location, hashtag, language, and country code. Then we label the tweets using their categories. For example, all tweets of Rockport area can be labeled with number of 4, which was the hurricane category of Rockport. In our research, we use labels "12" to annotate the tweets from hurricane category 1 and 2 areas, and use "34" to annotate the tweets from hurricane category 3 and 4 areas due to the limited amount of geo-tagged tweets for those relevant areas.
*Generating Vector Model*: We use Word2Vec and Bag-of-Words to generate vectors for the data set. Bag-of-Words represents a text as a bag (multiset) of its words considering multiplicity, whereas Word2Vec model represents words in a continuous vector space.
*Classification and Comparison*: In this step, we apply different classification algorithms to the training sets obtained from the last step and compare the F1 score (a measure of a test's accuracy) to find the best prediction approach. A 10-fold cross validation method is used to test the models in the training phase in order to give an insight on how the model will generalize to a new dataset.
*Applying Prediction for a Location*: We choose two new places, which are not included in the training data sets. We predict each of the tweets and label them as "12" (Hurricane Level 1 or 2) or "34" (Hurricane Level 3 or 4). If there are more tweets of "12" than "34", we identify the place as category "12". In this way, we can predict the hurricane category for a specific location.

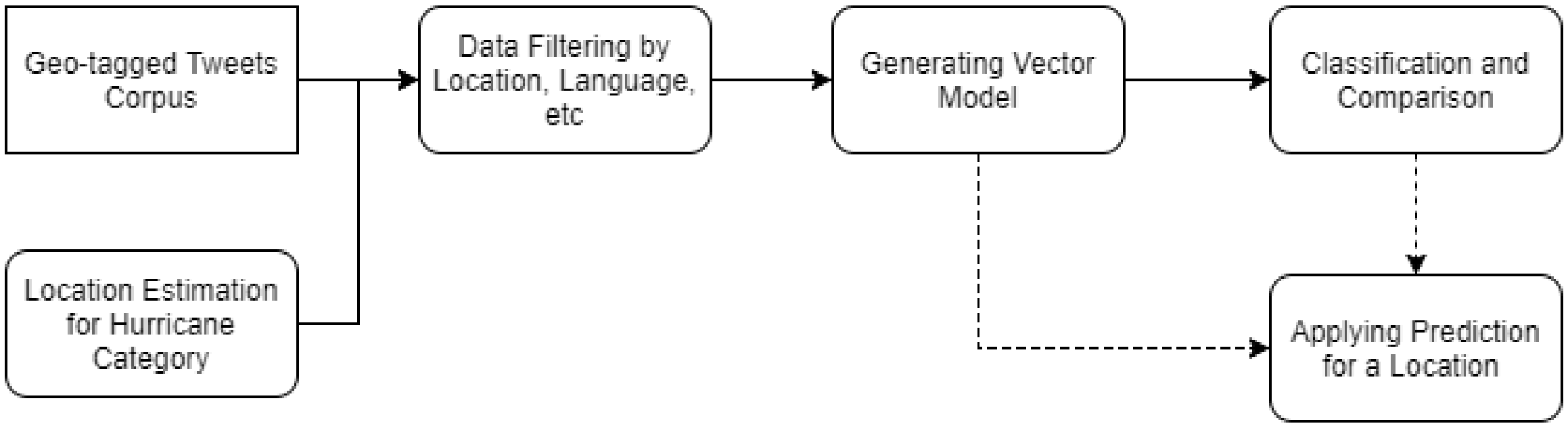

**Figure 1. Overview of Our Working Processes**

**Environmental Issues Setup**

Depending on the information given by the wunderground website, we can mark in the map using the latitude and longitude coordinates and the locations' related hurricane information. In this step, we can find the related towns and cities for each hurricane category. Figure 2 shows an example of a Twitter user post: "Rockport getting bad & it's just the start of the storm, don't underestimate #Harvey" at "8/25/2017 9:15:33 PM".

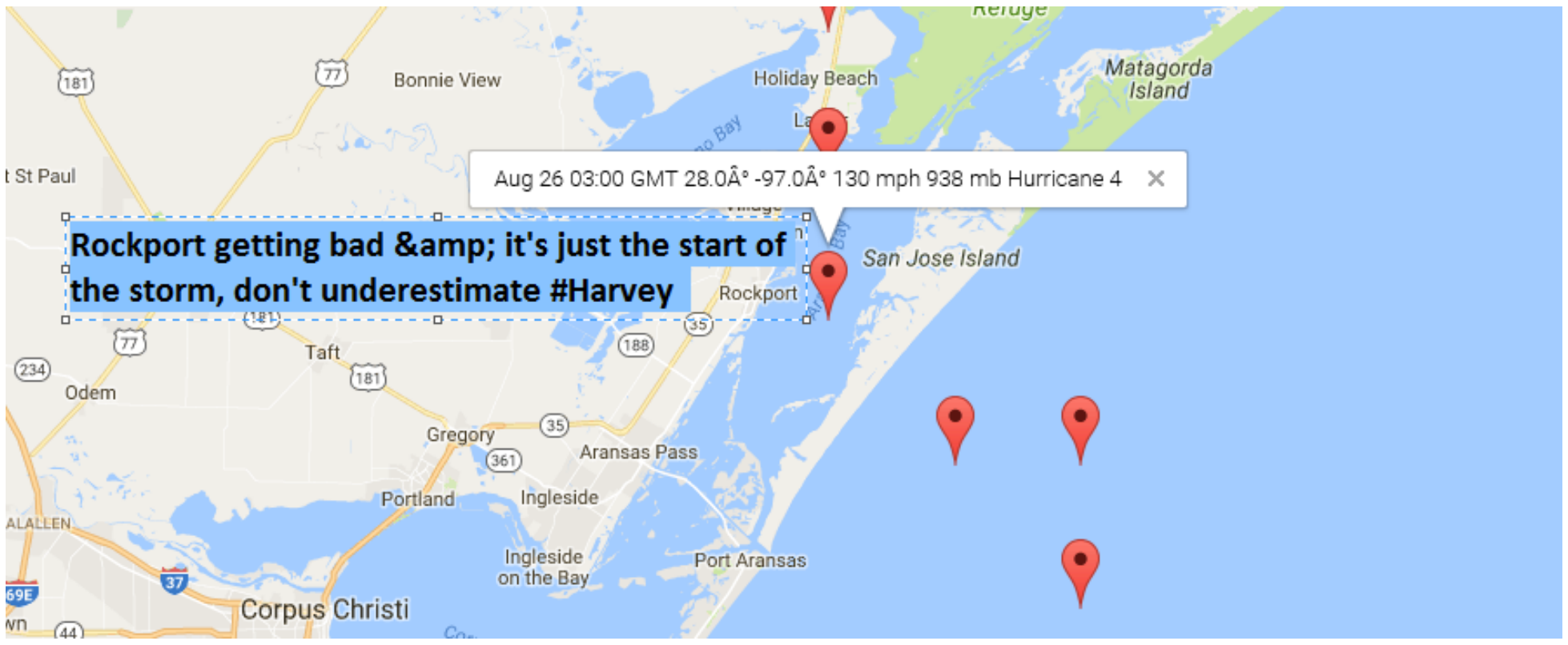

**Figure 2. A Tweet from Rockport of Texas.**

We first mark the places in the map with their category level and timing depending on the data from a trustworthy website (see Figure 3), and then identify the related places for a specific hurricane category. For example, for hurricane level 4, we identified Aransas Pass, Corpus Christi, Fulton, Holiday Beach, Port Aransas and Rockport. For each of them, we can filter their related tweets using hashtag "Harvey".

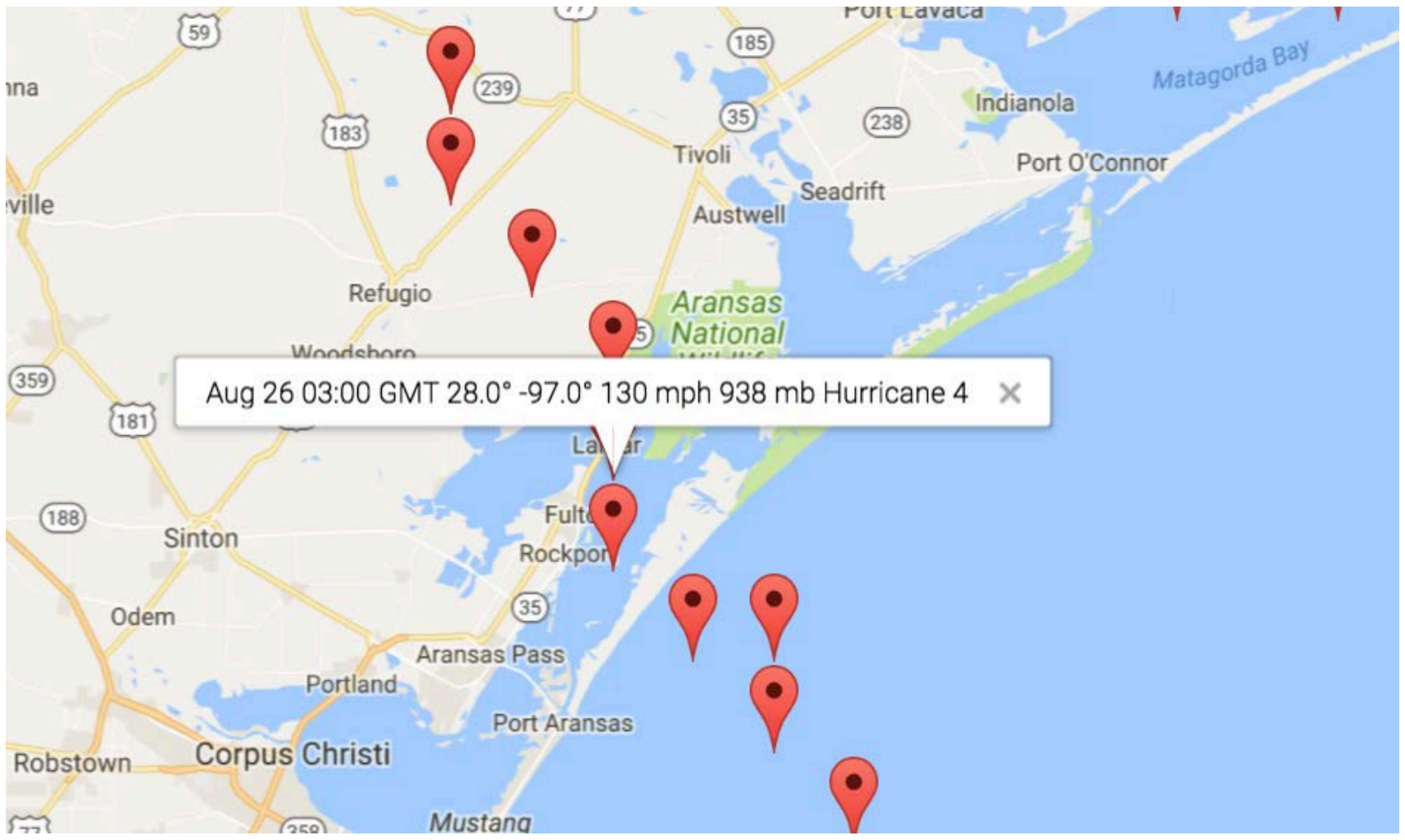

**Figure 3. Google Map Marker for Harvey**

### Twitter Data Corpus

The Twitter data set we use is a one-month (from August 17 to September 16) tweets set of Hurricane Harvey, Irma and Maria, and has geo location information. The total number of the tweets is 188,721. From the data set, we can filter the tweets by location and hashtags. By using a text and Twitter analytics tool called DiscoverText, we can easily preprocess the tweets for experiments.

### Data Filtering

In the Twitter data corpus, only a small part of the tweets has geo-coordinates, and all of them have location coordinates, which means we can search using location name – city name or town name. Although a lot of points are marked in the map, most of them are far from big cities, such as Houston or Miami, and highways. This leads to the conclusion that only hundreds of tweets are left for each category and its related locations. Table 1 shows an example of locations and number of tweets information for Harvey category 4 areas. In the future, if a new hurricane happens more data can be collected and combined with the historical data, but for this experiment, we only carry our experiment based on the available data sets.

**Table 1. Example Locations and Number of Tweets for Harvey Category 4 Area**

| Places | Level | Number |
|---|---|---|
| Aransas Pass | 4 | 5 |
| Corpus Christi | 4 | 416 |
| Fulton | 4 | 12 |
| Holiday Beach | 4 | 1 |
| Port Aransas | 4 | 13 |
| Rockport | 4 | 8 |

Because of the limited number of tweets, we put category 1 and 2 together to form a new label "12". Similarly, we put 3 and 4 together to have a new "34" label. In the labeling step, we label the tweets with "12" or "34". If we have more data in the future, we can split the categories into independent states for detailed analysis. For other filter methods, we use hurricane names like "Harvey" or "Irma". We only consider the tweets posted in English and inside the United States. The total numbers of tweets for each hurricane and category is shown in Table 2.

**Table 2. Tweets after Filtering with Location Name, Hashtag, Language and Country**

| | Category 12 | Category 34 |
|---|---|---|
| Harvey | 103 | 280 |
| Irma | 176 | 132 |

### Classification and Validation

To prepare the data for classification, we annotate the tweets of categories 1 and 2 as "12" and label the tweets of categories 3 and 4 as "34". Using this labeling method, we generate two original ground truth data, one for Harvey and one for Irma. Then we apply two different base line methods to build vectors for each of the ground truth data sets: Word2Vec and Bag-of-Words (BOW). For the Word2Vector method, we use Google's pre-trained Word2Vec model, which includes vectors for a vocabulary of 3 million words and phrases trained on part of Google News dataset (about 100 billion words). For the BOW model, we use the most common type of characteristics, which is term frequency, the number of times a term appears in the text. The experiment is performed to primarily verify the hypothesis that there exists a relationship between the hurricane category of an area and the tweets for that area using the corpus of Twitter data. Additional types of characteristics will be applied in future work.

We apply a two-step experiment to verify our hypothesis. After generating the training sets, we apply cross validation to the two training sets and a combination using 5 different classification algorithms (Decision Tree, Naïve Bayes, Logistic Regression, Random Forest and Support Vector Machines). After evaluating the results, we find the best feature generation methods and algorithms, and then use them in the second step of predicting two small towns' hurricane category for further analysis.

## EXPERIMENTAL RESULTS

Our two-step experimental results imply a positive relationship between the tweets posted in an area and the actual hurricane category in the area. We first use a 10-fold cross validation to validate each unique classifier model, and then apply a 10-fold cross validation to the combination of the two classifier models. Through analyzing the cross validation results we select the best feature generation baseline method and classification algorithm to predict two towns' hurricane categories.

### Cross Validation

We perform 10-fold cross validation on Harvey's data, Irma's data and the combined data of Harvey and Irma using different classification algorithms, namely Decision Tree (J48), Naïve Bayes, Logistic Regression, Random Forest and Support Vector Machines (SVM). These algorithms represent different categories of classification algorithms. We apply the five algorithms to find the one with the best performance so that we can use it to make prediction for small towns in the next step.

The cross validation results are plotted as shown in figures 4 and 5. F1 score is a harmonic average of the precision and recall to measure a test's accuracy, and it is used as the measurement for demonstrating the results in our experiments. Figure 4 presents the cross validation F1 results using Word2Vec as the base line method. We can see that Random Forest has the best performance for cross validation on Harvey's data and the value is 0.775. For cross validation on Irma's data, SVM has the highest score, which is 0.649. For cross validation on the combination of Harvey and Irma's data, Random Forest has the highest score, which is 0.657. From the results we can find a moderate positive relationship between the tweets posted in an area or several areas and the hurricane categories in those areas.

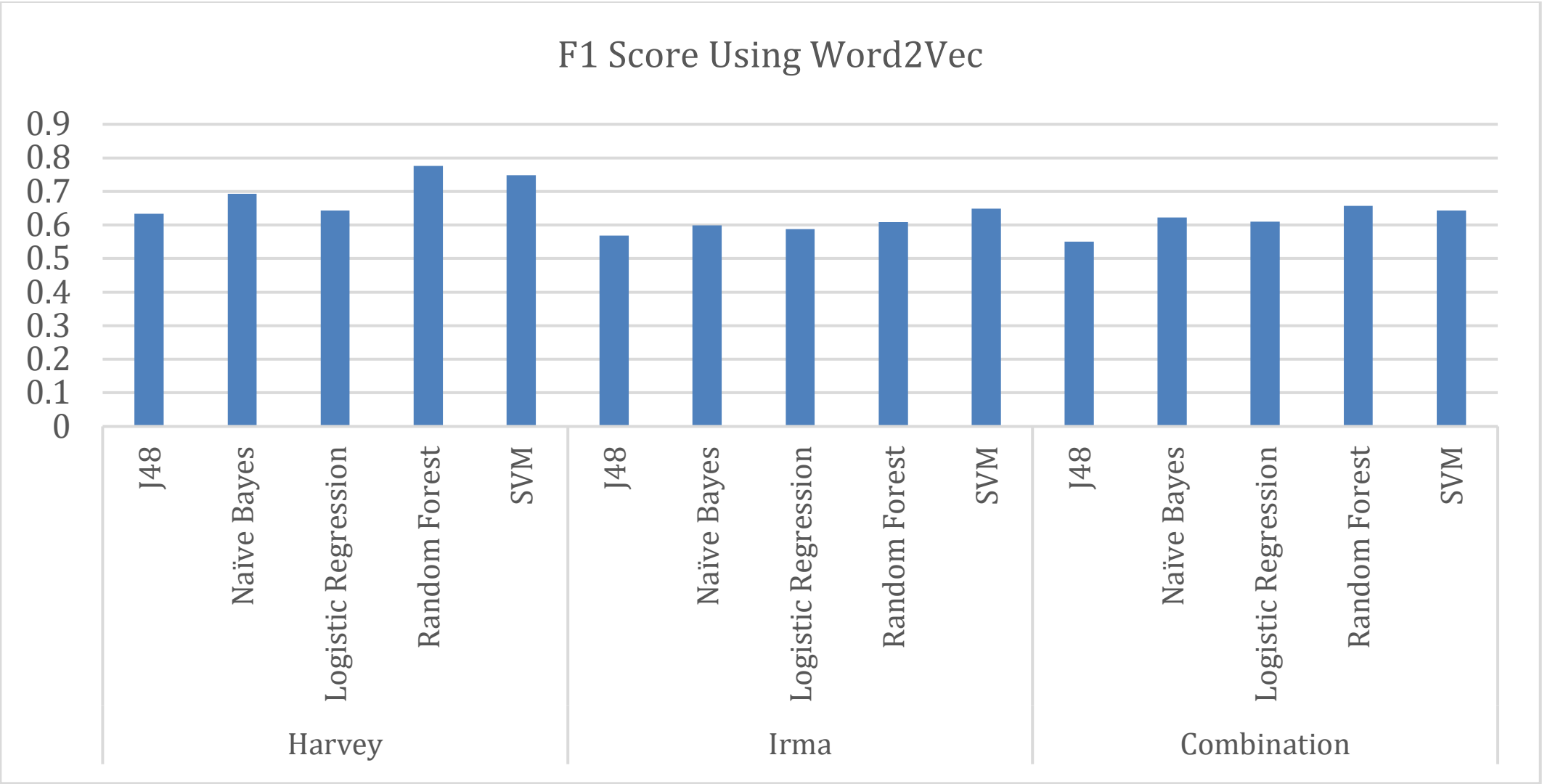

**Figure 4. Cross Validation F1 Results Using Word2Vec**

Figure 5 presents the cross validation F1 results using BOW as the base line method. We can see that Random Forest has the best performance for cross validation on Harvey's data and the F1 score is 0.796. For cross validation on Irma's data, Naïve Bayes has the highest score, which is 0.740. For cross validation on the combination of Harvey and Irma's data, Logistic Regression has the highest score, which is 0.829. Depending on the comparison of all the F1 scores presented in the two figures, we find Bag-Of-Words has an overall better performance

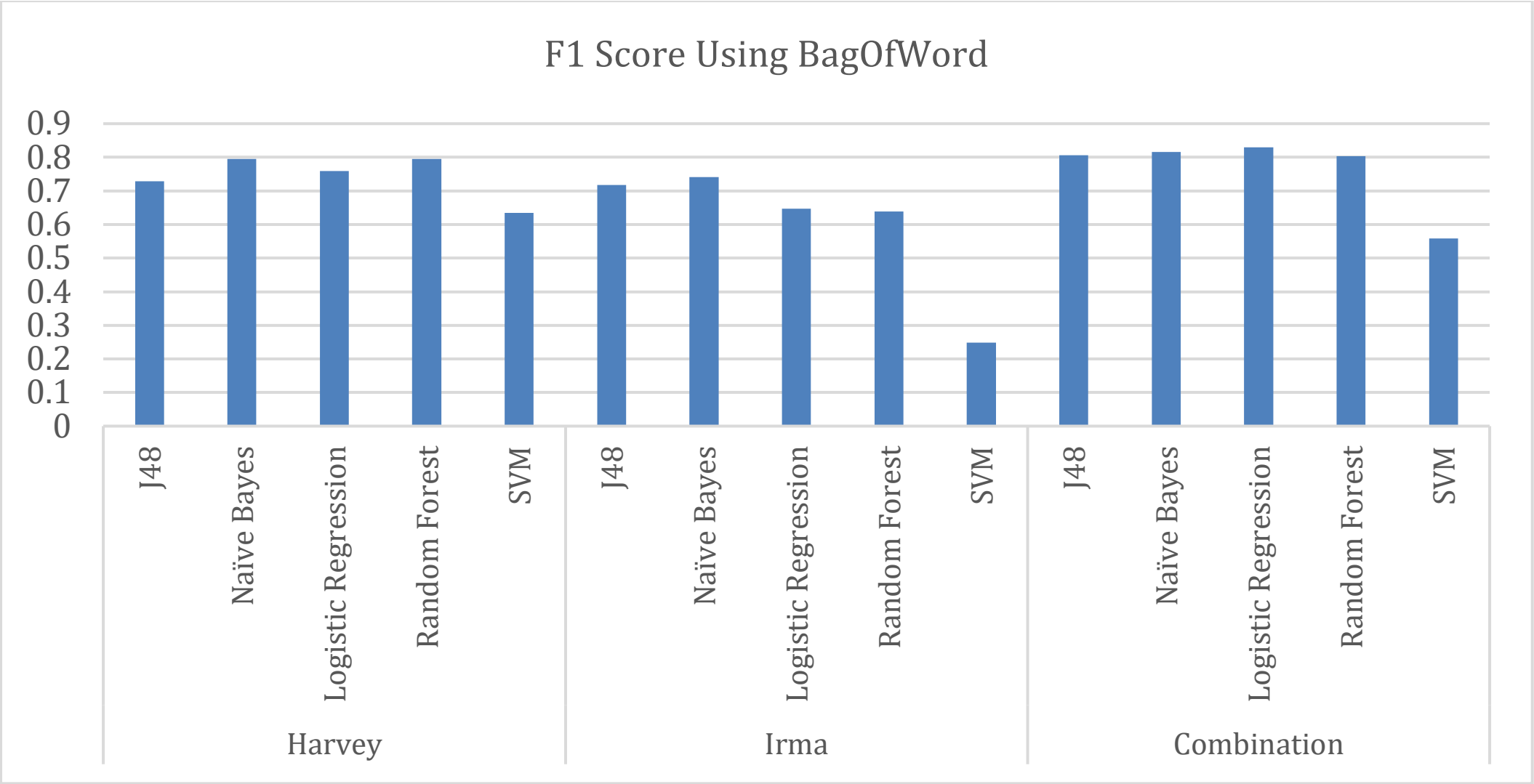

**Figure 5. Cross Validation F1 Results Using Bag of Word**

In the next step of experiment, we choose Random Forest algorithm to apply on Harvey training set, Naïve Bayes for Irma training set, Logistic Regression for a Combinational training set. Bag-of-Words will be used as the method for feature generation.

### For Validation of a Selected Area

A positive relationship between Twitter data and hurricane category can be recognized through the result of the last step of the experiment. We use three different training sets to predict two small towns' hurricane category, namely, Plant City, Florida and Crystal River, Florida. Both towns were impacted by hurricane Irma. By classifying the tweets for the towns and applying a proposed predicting method, we have successfully predicted the categories for the two towns. Thus, the proof of our initial hypothesis is further enhanced.

For predicting Plant City, Florida, by using Irma training set and the Naïve Bayes algorithm, we can correctly predict 89.2% instances of tweets; by using Harvey training set and the Random Forest algorithm, we can correctly predict 2.7% instances; by using a combination training set and the Logistic Regression algorithm, we can correctly predict 86.5% instances. For predicting Crystal River, Florida, by using Irma training set and the Naïve Bayes algorithm, we can correctly predict 83.3% instances; by using Harvey training set and the Random Forest algorithm, we can correctly predict 0% instances; by using a combination training set and the Logistic Regression algorithm, we can correctly predict 58.3% instances. The results are shown in Figure 6.

Through the low prediction precision of using the Harvey training set and the Random Forest method to predict the two small towns' hurricane categories, we can see that it is not fit to use one hurricane's training set to predict a place in another hurricane's area. However, it is possible to use the combination of the training sets to do the prediction. The results imply that using related hurricane data to predict for related areas is the best approach. The reason may be that the posts in one hurricane's area consist of area-specific or hurricane-specific words. For example, tweets about Hurricane Harvey usually contain words like "Texas", "Houston" and "Harvey". If we filter out the area-specific words and special names, the results of using one hurricane's Twitter data to predict another might be improved.

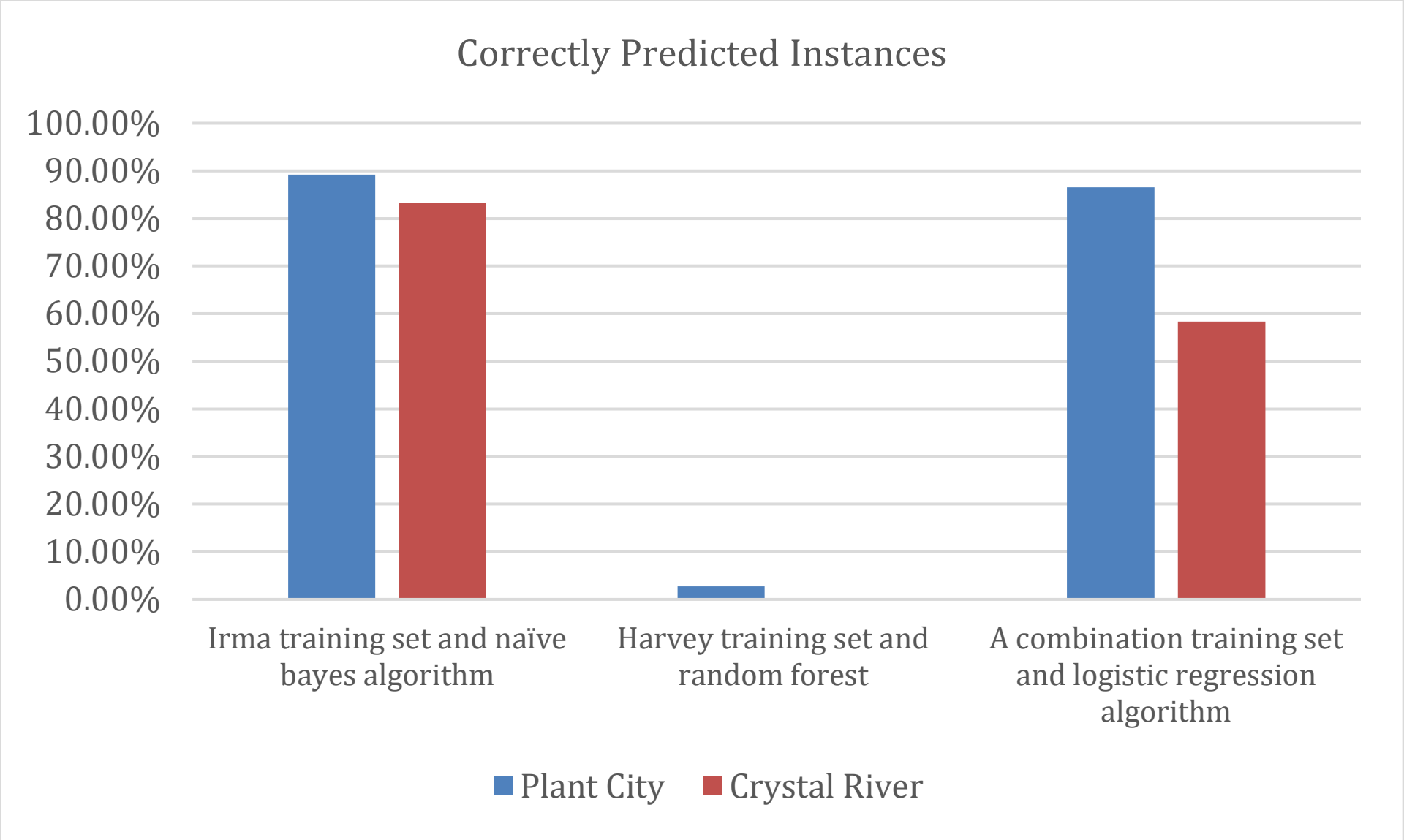

**Figure 6. Correctly Predicted Instances Using Different Training Sets and Algorithms**

To predict the hurricane level of a specific place, we propose an algorithm based on the observed results. The general steps are listed below. As described above, if we apply this algorithm, and use the Irma training set and the Naïve Bayes algorithm, or use a combination training set and the Logistic Regression algorithm, we can successfully predict the categories for the two places. The steps in the algorithm are:

1. Choose a proper training set and the most efficient classification algorithm.
2. Collect related tweets (set *s*) for a location.
3. Run classification algorithm to label each tweet in *s*.
4. Calculate the total number of tweets for each label, and find the majority label, which is of the final predicted category for that location.

Although our method of predicting hurricane categories performs well on the locations we chose, it can be not true for other locations. Different location can have different temporal, spatial or social features, which may affect the application of the method. It still needs more research based on more real world data to prove the efficiency of the method, as well as to improve the method.

## DISCUSSION AND FUTURE WORK

To the best of our knowledge, this paper is the first attempt to use social media data to analyze and predict the category of a hurricane, or even the severity of a disaster. We can continuously collect the Twitter data for areas affected by hurricanes and verify the idea using the methods presented in this paper and a larger corpus of data. It is possible to analyze the relation between a disaster like hurricane and related social media data due to the wide area a hurricane can affect. However, during our experiment, we found several issues, other than the limit of data corpus, which necessitate further investigation as part of the future work.

*Reasons about the Relation*: It would be interesting to analyze the root cause of why there is a relationship between tweets and hurricane level. One possible approach could be using Bag-of-Words to find the most common words to make comparison.

*Hurricane-specific Data*: The precision is low if we use Harvey data to predict Irma data with respect to hurricane level. However, the precision is more significant if we use cross validation within Harvey data or Irma data. We found a lot of hurricane-specific words like “Florida”, “Key West”, “Texas”, and “Rockport”. For locations, they are unique to a specific area, making it hard to use one hurricane’s data to predict another hurricane’s severity unless we remove these words. This will be explored in our future work. If the hurricane-specific words or some irrelevant words are deleted from the text, the accuracy may be improved.

*Temporal Factor*: We are using one month of Twitter data. Another approach would be to use data in a specific period like the exact day the hurricane occurs. There would be less data if the period is short, however, since hurricanes usually happen in the same places (coastal areas), we can accumulate data of each hurricane to build a larger corpus as the ground truth. The tweets can be further categorized into three sets: before, during and after the hurricane. More complex relations can be used as a pattern to identify the category of the hurricane.

*Number of Tweets*: The tweets for each category we use in our experiment are not evenly distributed. Although the usage of vector generation methods can ease the effects of corpus size differences, the corpus size can impact the prediction results. In future experiments, we can take the number of tweets into consideration and examine what kind of impact it may bring to the experiment.

*Threshold of Prediction*: For predicting the category of a hurricane in a specific area, we can use a small set of tweets or a large set of tweets. The exact number depends on how many tweets can be obtained from the source data corpus. Future work needs to identify the exact threshold of the number of tweets, which can make the prediction most significant. The threshold may be determined by historic data or may also be determined by current hurricane related data.

*Human Statistical Factor*: This study did not consider the credibility of the authors of the tweets. This may be a limitation of our paper. The tweets can be posted by different groups: namely, individuals, newspapers, government departments, or advertising agents. Each group may have different demographic and ideological impacts on the experimental results.

The application of disaster prediction can help in real-time disaster response and decision making for governmental or civil agents. For a specific area experiencing a hurricane, we can predict the hurricane category using real-time Twitter data for that area by using historic data. It can be more meaningful if we apply our method to an area where it is hard to determine the severity of the damage, such as a valley area, high mountain areas, or near-coastal areas. Governmental or civil agents can use the evaluation results of social media data analysis as a reference. This paper explores hurricane categories, other disaster types might benefit from similar research methods.

## CONCLUSION

In this paper, we introduced a study of finding correlation between Twitter data and hurricane category. Methods used and the preliminary results were presented. A moderate positive relationship was observed based on the experimental results. The contribution also includes a release of the data set, which was generated over the course of the study. We present an algorithm and implementation results of predicting an area’s hurricane category. Depending on the experimental results, we can successfully predict the hurricane category for a given location.

However, intuitively, if we only have a small set of tweets for a small town, it is hard to predict the hurricane category. For future work, we can determine the proper magnitude of an area or the threshold number of tweets, related to the area for yielding better prediction results.

## ACKNOWLEDGEMENTS

Thanks to Dr. Jacob Groshek, Boston College, for his grateful help with an introduction to Dr. Stuart Shulman. Dr. Shulman provides a tool called DiscoverText designed for big data processing and applying different learning techniques. Our research would not have been accomplished without their help in providing the related research data.